\documentclass[aps,prl,superscriptaddress]{revtex4-2}
\usepackage{graphicx}
\usepackage{xcolor}
\usepackage{soul}
\usepackage{lipsum}

\begin{document}

\title{Mean-field behavior of the finite size Ising model near its critical point}

\author{D. Olascoaga-Rodr\'{\i}guez}
\affiliation{Departamento de F\'{\i}sica, Universidad Aut\'onoma Metropolitana - Iztapalapa, \\
Av. San Rafael Atlixco No. 186, Col. Vicentina\\
09340 Ciudad de M\'exico, Mexico}
\author{F. Sastre}
\affiliation{Departamento de Ingenier\'ia F\'isica,\ Divisi\'on de Ciencias e Ingenier\'ias,
Campus Le\'on de la Universidad de Guanajuato,
AP E-143, 37150 Le\'on, Guanajuato, Mexico}
\author{V. Romero-Roch\'{\i}n}
\email{{\rm Contact author:} romero@fisica.unam.mx}
\affiliation{Instituto de F\'{\i}sica, Universidad Nacional Aut\'onoma de M\'exico, \\ 04510 Ciudad de M\'exico, Mexico}

\date{\today}

\begin{abstract}
Universality classes encompass the analogous thermodynamic behavior of unlike physical systems, at different spatial dimensions $d$, in the vicinity of their critical point. Critical exponents define these classes, with the Ising model being the outstanding prototype that elucidates the differences from the mean-field category, believed to be valid above a critical dimension only. 
Here, in apparent striking contradiction to the Ising universality class, we demonstrate that the critical behavior of a {\it finite} Ising system of $N$ spins in $d = 3$  obeys mean-field Landau theory in the vicinity of {\it its} critical point, with classical critical exponents. Yet, when expressed in terms of the linear size $L$ of the system, the free energy unveils its proper finite-size scaling form, from which the thermodynamic limit critical temperature $T_c$ and the Ising critical exponents $\nu$, $\gamma$ and $\beta$ can be identified. We find that the larger the size $L$, the smaller the mean-field region, shrinking to zero in the thermodynamic limit.
These conclusions are achieved via the use of an alternative approach to collect data from a Monte Carlo simulation of a three-dimensional Ising model that allows for the evaluation of the 
free energy per spin $f = f(T,m;L)$ and of the coexistence curve, or spontaneous magnetization at zero magnetic field, $m_{\rm coex} = m(T;L)$ as  functions of temperature $T$ and magnetization per spin $m = M/N$. Our results suggest a revision of the role of mean-field theory in the elucidation of critical phenomena.
\end{abstract}

\maketitle

{\it Introduction--}The Ising model\cite{Ising1925,Onsager1944,Yang1952,Brush} is the paradigmatic system for our understanding of critical phenomena in continuous phase transitions. It is well understood that in one dimension, $d = 1$, there is no transition, while for $d \ge 2$ one obtains critical exponents belonging to their proper universality classes, rigorously validated by the Renormalization Group (RG) \cite{Kadanoff,Wilson,FisherRMF,Ma,Amit,Kardar}. Furthermore, in $d = 2$ one knows the exact exponents via Onsager \cite{Onsager1944} and Yang \cite{Yang1952} calculations. In $d = 3$, while there are no exact solutions, a very long history of high temperature expansions \cite{Essam:1968aa,Tarko:1975aa,Liu:1989aa,Butera:2002aa}, RG calculations \cite{Wilson,Kadanoff,Ma,Amit,Kardar}, field theory \cite{Chang:2025aa}, and Monte-Carlo (MC) simulations by means of the finite-size scaling procedure \cite{Fisher1972,Landau,Suzuki,Barber1985,Wolff,Ferrenberg1991,Ferrenberg1991b,Hilfer,Gupta,Ito,Blote1996,BinderPR,Hasenbuch2001,Hasenbuch2010,Lundow2009,Ferrenberg2018}, have lead to a robust consensus of the critical exponents values.  Moreover, RG has also shown that for $d \ge 4$ the critical exponents are those of Landau mean field theory, maintaining that fluctuations play a negligible role in determining critical behavior. Landau theory, remarkably clear and simple, sufficing to unravel the thermodynamic origin of the transition, can also be established, as Landau himself did it  \cite{Huang,Ma,LL,Kardar}, by considering a truncated Taylor series expansion of the free energy around the assumed critical point, with the suitable symmetry properties of a magnetic material, but without ever including the dimensionality nor explicitly taking the thermodynamic limit,
\begin{equation}
f(m,T) \approx f_0(T) + a_0 (T-T_c) m^2 + b m^4 \>. \label{LandauF}
\end{equation}
The Landau coefficients $a_0 > 0$ and $b > 0$ are system parameters, and the critical point is $m = 0$ and $T = T_c$. Even without the knowledge of $f_0(T)$, one obtains the classical or mean-field critical exponents $\beta_L = 1/2$, $\gamma_L = 1$ and $\delta_L = 3$ \cite{Huang,Ma,Kardar}. In addition, this also leads to the well known result that, if one {\it assumes} that a free energy is an analytic function at the critical point, then Landau theory follows and, therefore, the classical critical exponents ensue. The analytic van der Waals model \cite{LL,Stanley} of a liquid-gas phase transition is in the same category with the same critical exponents. In both models the specific heat does not diverge and, therefore, one considers $\alpha_L = 0$. It goes without say that this set of exponents obeys Widom scaling laws \cite{FisherRMF,Widom1965}.\\

In this paper, by means of a different way to collect data in a usual Monte Carlo simulation, and that does not require to perform averages, we are able to extract the free energy per spin $f(m,T;L)$, which depends on the finite system size $L$. The unmistakable conclusion is that, for a {\it finite} system,  $f(m,T;L)$ has the Landau form (\ref{LandauF}) but with parameters that depend on $L$. Remarkably, at the same time, the finite-size scaling expression of the free energy \cite{Fisher1972,Landau,Suzuki,Barber1985} naturally emerges when the $L$-dependence of the coefficients is explicitly considered,
%At the same time, and as a direct consequence of our calculations for different values of $L$, one finds that the mean-field free energy 
%%{\color{red} transforms itself into the expected finite-size scaling form}  
%can be expressed in the expected finite-size scaling form \cite{Fisher1972,Landau,Suzuki,Barber1985},
thus allowing for the identification of the well known 3D thermodynamic limit critical temperature $T_c$ and the Ising exponents $\nu$, $\gamma$ and $\beta$ \cite{Butera:2002aa,Deng:2003aa,Lundow:2010aa,Ferrenberg2018,Chang:2025aa}. 
That is, while the finite-size free energy is mean field around its own finite-size critical point, yielding classical exponents, its parameters bear the thermodynamic limit exponents and critical temperature in the finite-size scaling form. 
In addition, we are able to extract the full coexistence curve or spontaneous magnetization and, again, near its critical temperature the behavior is mean-field, while quickly approaching the thermodynamic limit values far from the critical region, even for relative small values of $L$. The evident further conclusion is that the small critical mean-field region shrinks to zero as $L$ grows with no bound, thus allowing for the full emergence of the expected Ising non-analytic behavior in such a limit. The description of this approach to the thermodynamic limit is not in accord with the usual explanations of critical phenomena, suggesting that the validity of mean-field theory should be brought back to scrutiny. We believe these observations open the door for a, yet, another aspect of our understanding of critical phenomena since, after all, real systems are finite.\\

{\it A method for calculating the free energy--}Let us first recall that for any spin configuration or state of the system, $\{s\} = (s_1, s_2, \dots, s_N)$, with $s_i = \pm 1$,  its energy and magnetization, in reduced units, are $E_{\{s\}} = -\sum_{<ij>} s_i s_j$, summing over nearest neighbors in a cubic lattice with periodic boundary conditions, and $M_{\{s\}} = \sum_{i=1}^N s_i$. Then, in the ensemble $(N,M,T)$, the partition function is
\begin{equation}
Z(N,M,T) = \sum_{E} \Omega(E,M) \> e^{-E/T} \>,\label{Z}
\end{equation}
where $\Omega(E,M)$ is the number of spin configurations with energy $E_{\{s\}} = E$ and magnetization $M_{\{s\}} = M$. The associated free energy is $F(N,M,T) = - T \ln Z(N,M,T)$. The entropy $S$ and the magnetic field $H$ follow,
\begin{equation}
S = - \left(\frac{\partial F}{\partial T}\right)_{N,M} \>\>\>\>\>\>\>\> H = \left(\frac{\partial F}{\partial M}\right)_{N,T} \>.\label{SH}
\end{equation}
It is important to highlight that this is {\it not} the usual ensemble typically used in most of the articles and textbooks studies on the Ising model. That one is $\tilde Z(N,H,T)= \sum_{\{s\}} e^{-(E_{\{s\}} - H M_{\{s\}})/T}$, where the sum is over {\it all} spin configurations.  The relationship between the ensembles is $\tilde Z(N,H,T) = \sum_M Z(N,M,T) e^{HM/T}$, becoming fully equivalent in the thermodynamic limit.\\

 Here we calculate a {\it fraction} of the partition function $Z(N,M,T)$, for given $N$ and $T$, as a function of $M$. 
We generate a very large sequence  of ${\cal N}$ spin configurations by means of Monte Carlo spin-flip {\it trials}. Let ${\cal Z}_{\cal N}(N,T,M)$ be the number of configurations in a sequence with magnetization $M$. It is calculated as follows. Let $\{s\}$ be the current configuration in a sequence, after a given number of steps, with $M = M_{\{s\}}$ its magnetization. Now perform a single random spin flip, yielding a new configuration $\{s\}^\prime$. Apply the usual Metropolis step of comparing a random number $0 < {\cal R} < 1$ with ${\rm exp}[-(E_{\{s\}^\prime}-E_{\{s\}})/T]$: if accepted, the new configuration is $\{s\}^\prime$; if not, the new configuration is {\it again} the current one $\{s\}$.  Then, update ${\cal Z}_{\cal N}(N,T,M) = {\cal Z}_{\cal N}(N,T,M) + 1$. Repeat the cycle  ${\cal N}$ times. Importantly, all configurations in the sequence are counted, some of them more than once if the spin-flip is not accepted. But if the spin-flip is accepted the magnetization changes and, therefore, we end up finding ${\cal Z}_{\cal N}(N,T,M)$ for all values of allowed $M$, if ${\cal N}$ is large enough.  The output is that ${\cal Z}_{\cal N}(N,T,M)$ is a fraction of $Z(N,M,T)$. As it is known \cite{Metropolis,LandauBinder}, Metropolis method samples spin configurations  $\{s\}$ with relative probability $e^{-E_{\{s\}}/T}$. On the other hand, all configurations with the same energy $E$ and magnetization $M$ are equally probable, with microcanonical probability $1/\Omega(E,M)$. Hence, by summing over all states with the same magnetization $M$, already distributed by $e^{-E/T}$, we are counting the states that contribute to the sum in (\ref{Z}), with their respective relative probability. 
The crucial step of the method, where the current state is counted again if the Monte Carlo step is not accepted, is based on the mentioned fact that all states with the same $E$ and $M$ are equally probable.\\

To be more precise, with no rigorous proof, we claim that
\begin{equation}
\lim_{{\cal N} \to \infty} \frac{{\cal Z}_{\cal N}(N,T,M)}{{\cal Z}_{\cal N}(N,T,M^\prime)} = \frac{{Z}(N,T,M)}{{Z}(N,T,M^\prime)} \>.\label{Zrat}
\end{equation}
If true, this implies that ${\cal Z}_{\cal N}(N,T,M) \approx \zeta({\cal N},N,T) {Z}(N,T,M)$ with $\zeta({\cal N},N,T)$ an unknown number 
that does not depend on the magnetization and that should scale linearly with ${\cal N}$. The statement that the ratio ${\cal Z}_{\cal N}(N,T,M)/{\cal Z}_{\cal N}(N,T,M^\prime)$ is independent of ${\cal N}$ and depends on $M$ and $M^\prime$ only, for given $N$ and $T$, can be numerically verified, regardless of the meaning of ${\cal Z}_{\cal N}(N,T,M)$, as discussed in the Appendix. 
The usefulness of ${\cal Z}_{\cal N}(N,T,M)$, with the claim (\ref{Zrat}), resides on the observation that $\ln {\cal Z}_{\cal N}$, up to an additive unknown temperature dependent function, yields the free energy $F =- T \ln Z \approx -T \ln {\cal Z}_{\cal N}(N,M,T) - T \ln \zeta({\cal N},N,T)$. This allows for retrieving thermodynamic properties such as the magnetic field $H$, see (\ref{SH}) and the magnetic susceptibility $\chi = (\partial m/\partial H)_T$.\\

The present procedure to evaluate ${\cal Z}_{\cal N}(N,T,M)$ is a modification of the methods in Refs. \cite{Kastner:2000aa,Pleimling} and \cite{SastreIsing,SastreSW}, with some similar conclusions regarding the mean-field aspect for finite $L$ here discussed.  We insist that the method does not take the average of any quantity. In the Appendix
we show the values of ${\cal N}$ used for a given temperature $T$ and for the $L$ values here considered. We also mention that there have been similar analysis of the free energy and/or magnetization distribution function \cite{Hilfer,Binder1981,Eisenriegler,Valleau}, to those presented here; however, there was no connection to the mean-field results discussed below.\\

\begin{figure}[ht!]
\includegraphics[width=0.8\textwidth]{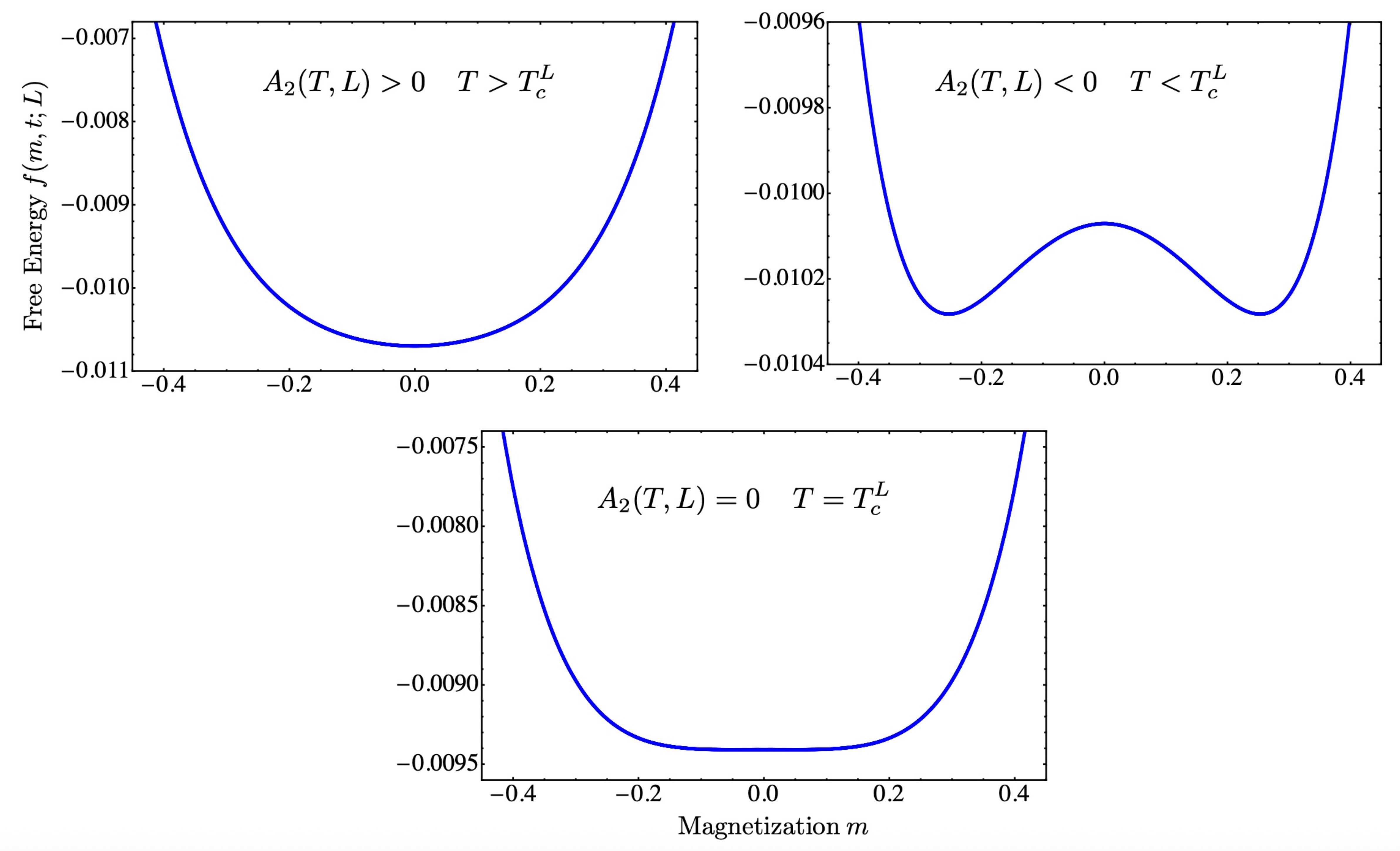}
\caption{Finite size free energy $f(m,T;L)$ as a function of magnetization $m$, above, at, and  below $T_c^L$, for $L = 22$. 
These curves are not fittings, appearing continuous due to the large number of magnetization points and the number of configurations, ${\cal N} \approx 5\times 10^{13}$ for each temperature}
\label{fvsm22}
\end{figure}

{\it Landau free energy and classical exponents for the finite-size Ising model--}Fig. \ref{fvsm22}, as a typical case, shows the finite free energy $f(m,T;L) = - (T/N)  \ln {\cal Z}(N,M,T;L)$ for $L = 22$, for three temperatures $T > T_c^L$, $T \approx T_c^L$ and $T < T_c^L$. Since one can quite accurately fit 
\begin{widetext}
\begin{equation}
f(m,T;L) = f_0(T,L) + A_2(T,L) m^2 + A_4(T,L) m^4 + A_6(T,L) m^6 +  A_8(T,L) m^8 + \dots \>, \label{ffit}
\end{equation}
\end{widetext}
 up to 10th order in our case, the finite-size critical temperature is unambiguously defined when $A_2(T_c^L, L) = 0$.  
The magnetic field $H(m,T;L)$, see (\ref{SH}), is a function of odd powers in $m$ and, evidently,
 at zero magnetic field, the free energy $f(m,T;L)$ shows a single minimum at $m = 0$ for $T \ge T_c^L$ and, for $T < T_c^L$, one observes two-phase coexistence, namely, there are two minima $m_{\rm coex}(T,L) = \pm |m(T,H=0;L)|$, yielding the coexistence curve or spontaneous magnetization, for the given value of $L$. 
 Although we can numerically find $f_0(T,L)$, it does depend on the unknown function $\zeta({\cal N},N,T)$ and, therefore, this prevents us from obtaining the entropy and the specific heat.
  
\begin{figure}[ht!]
\includegraphics[width=0.8\textwidth]{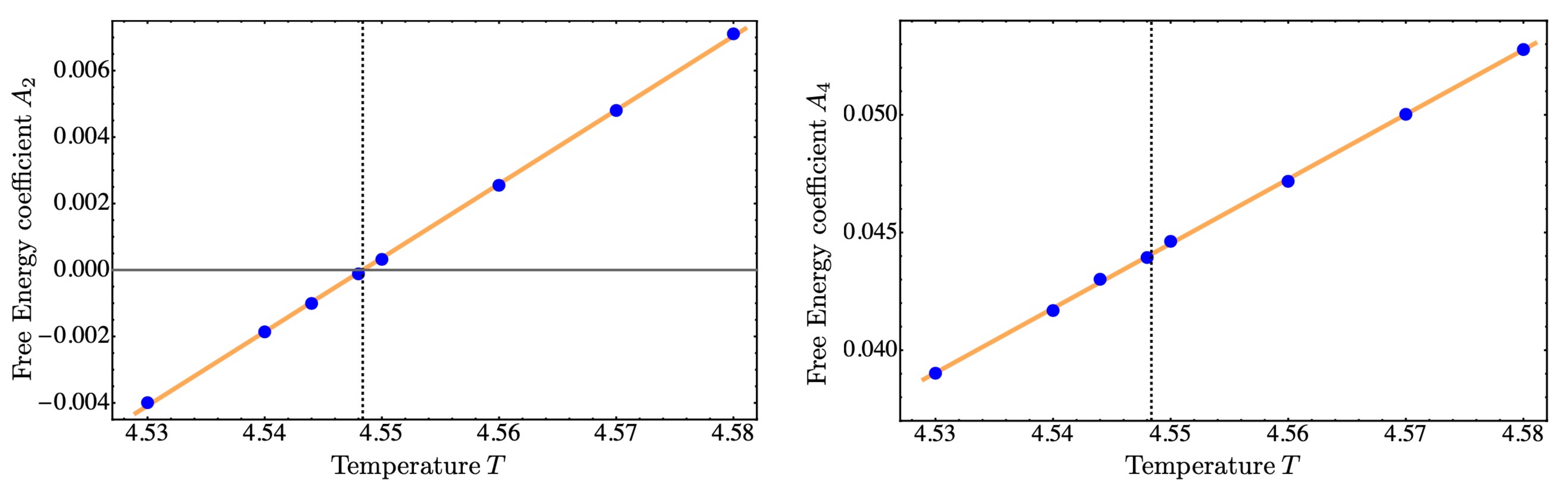}
\caption{Fitting coefficients $A_2(T,L)$ and $A_4(T,L)$ for $L = 22$, in the vicinity of the critical point $T_c^L$, see (\ref{ffit}). One finds, very accurately, $A_2(T,L) \approx a_0(L) (T - T_c^L)$, thus fitting $a_0(L)$ and $T_c^L$, the latter shown with a vertical dashed line in both plots. Then, we fit $A_4(T,L) \approx b(L) + b^\prime(L)(T-T_c^L)$. }
\label{AvsT}
\end{figure}

For temperatures $T$ near $T_c^L$ one can find the behavior of the coefficients $A_2(T;L)$, $A_4(T;L)$, $A_6(T;L)$ and so on in (\ref{ffit}) for a given $L$, as shown in Fig. \ref{AvsT}. Close to the critical temperature $T_c^L$ the dependence should be linear in $T$ for all coefficients, although the best fits are for $A_2$ and $A_4$. The coefficient $A_6$ is still very noisy for the typical configurations ${\cal N}$ here used. We fit $A_2(T,L) \approx a_0(L)(T-T_c^L)$ thus finding the critical temperature $T_c^L$ and the coefficient $a_0(L)$. Then, we fit the coefficient $A_4(T,L) = b(L) + b^\prime(L) (T-T_c^L)$. With these two approximations, typically very precise, and neglecting for the moment $b^\prime(L)$, one finds that, indeed, $f(m,T;L)$ approximates to the mean-field Landau expression (\ref{LandauF}), with $a_0$, $T_c$ and $b$ being functions of the system size $L$. 
%The same conclusion can be reached for the Ising model in 2D, not shown here. 
This behavior occurs for all values of $L$ that we have calculated, thus strongly evidencing that, near its critical point, $m = 0$ and $T=T_c^L$, the {\it finite} Ising model is mean-field, with classical critical exponents $\beta_L = 1/2$ and $\gamma_L = 1$.   It is interesting to point out that, in this ensemble, one calculates the {\it inverse} of the susceptibility $\chi^{-1} = (\partial H/\partial m)_T$, thus $\chi^{-1} \sim |T-T_c^L|^{\gamma_L} \to 0$, hence avoiding a diverging quantity in a finite system.
In the Appendix we show the values of the critical temperature $T_c^L$ and of the Landau coefficients $a_0(L)$ and $b(L)$ for the different $L$ cases here studied. \\

Our first simple explanation for this mean-field critical behavior is that, since the system is finite, the corresponding free energy can be analytically expanded around its critical point, thus unavoidably yielding Landau's prescription, but with true 3D Ising $L$-dependent coefficients. 
%We will return to the discussion of this result from a physical point of view.  
An important consideration is the estimation of next higher order terms in Landau free energy in order to assess its range of validity. One is the correction to $A_4(T,L)$ proportional to $b^\prime(L)$
and the other $A_6(T_c^L,L) \approx c(L)$. In the Appendix we show $b^\prime(L)$ as a function of $L$. Suffice to say here that this coefficient {\it grows} as $L$ increases, while $a_0(L)$ and $b(L)$ diminish. This indicates that the region of validity of mean-field Landau behavior becomes smaller for larger systems. We will go back to this point below.\\

\begin{figure}[ht!]
\includegraphics[width=0.5\textwidth]{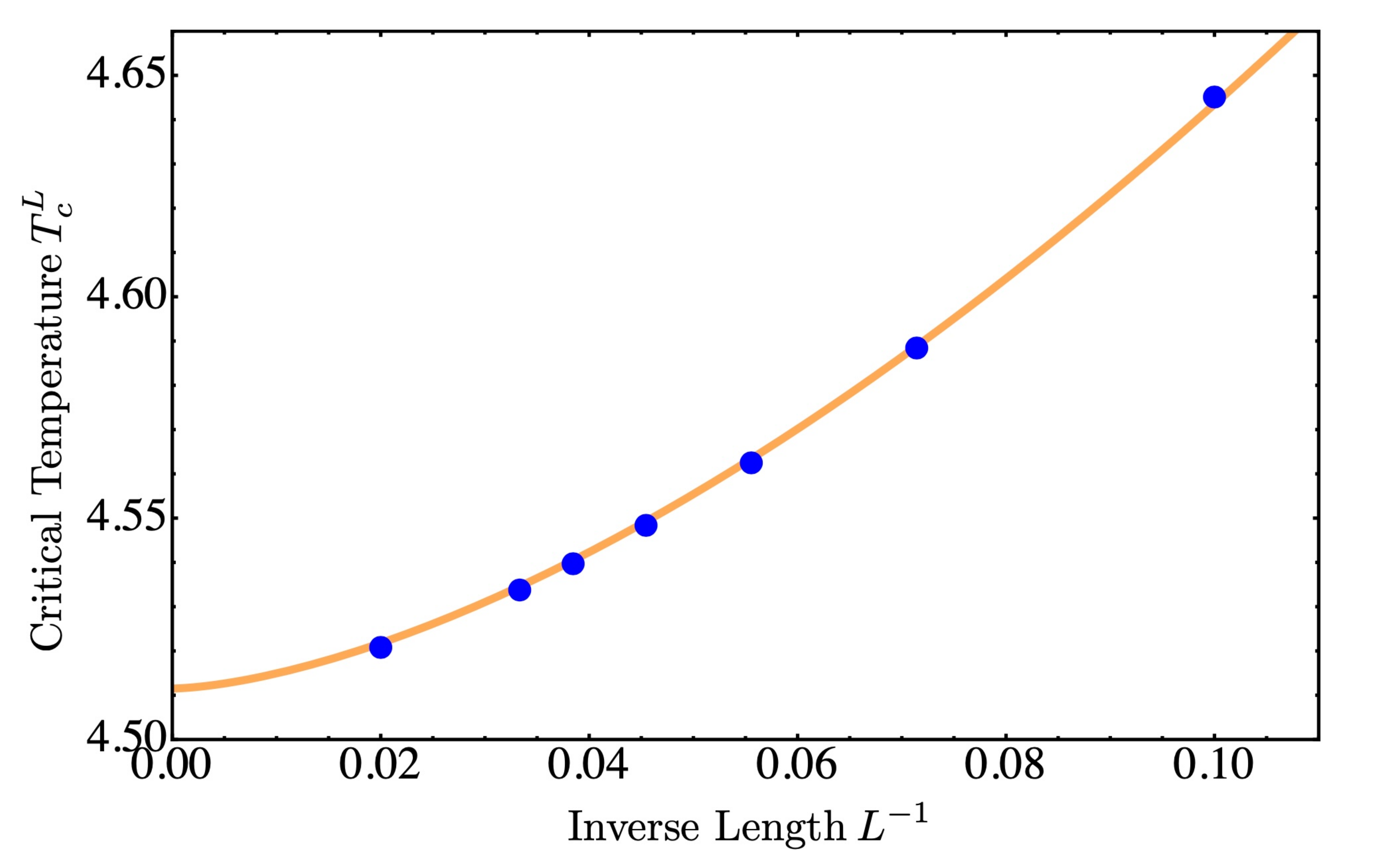}
\caption{Non linear fittings of $T_c^L  \approx T_c + {\cal L}_0 L^{-1/\nu}$ vs $L^{-1}$.  The dots are the results of the fittings for $L = 10, 14, 22, 26, 30, 50$ given in the Appendix.}
\label{FSS-fig-1}
\end{figure}

{\it Finite size scaling and the emergence of the Ising universality class--}We turn our attention to the limit $L \to \infty$. First, using finite-size scaling theory guidance \cite{Fisher1972,Landau,Suzuki,Barber1985}, we expect that 
$T_c^L  \approx T_c + {\cal L}_0 L^{-1/\nu}$, with $T_c$ and $\nu$ the thermodynamic limit critical temperature and exponent of the correlation length $\xi \sim |T-T_c|^{-\nu}$, and ${\cal L}_0$ a non-universal amplitude to be determined. 
Even with the small number of values of $L$ here considered, perhaps some of them not large enough, and given that there may be further scaling corrections \cite{Fisher1972,Suzuki,Barber1985,BinderPR}, we have performed a non-linear fitting \cite{Mathematica} of the above proposal obtaining $T_c = 4.51152(25)$, $\nu = 0.6297(9)$ and the coefficient  ${\cal L}_0 = 5.111(32)$. The values of $T_c$ and $\nu$ are well within the accepted values for the 3D Ising model \cite{Butera:2002aa,Deng:2003aa,Lundow:2010aa,Ferrenberg2018,Chang:2025aa}. Fig. \ref{FSS-fig-1} shows an excellent agreement of this fitting with our simulation values $T_c^L$. One can certainly achieve better accuracy with longer runs and larger values of $L$.

\begin{figure}[ht!]
\includegraphics[width=0.8\textwidth]{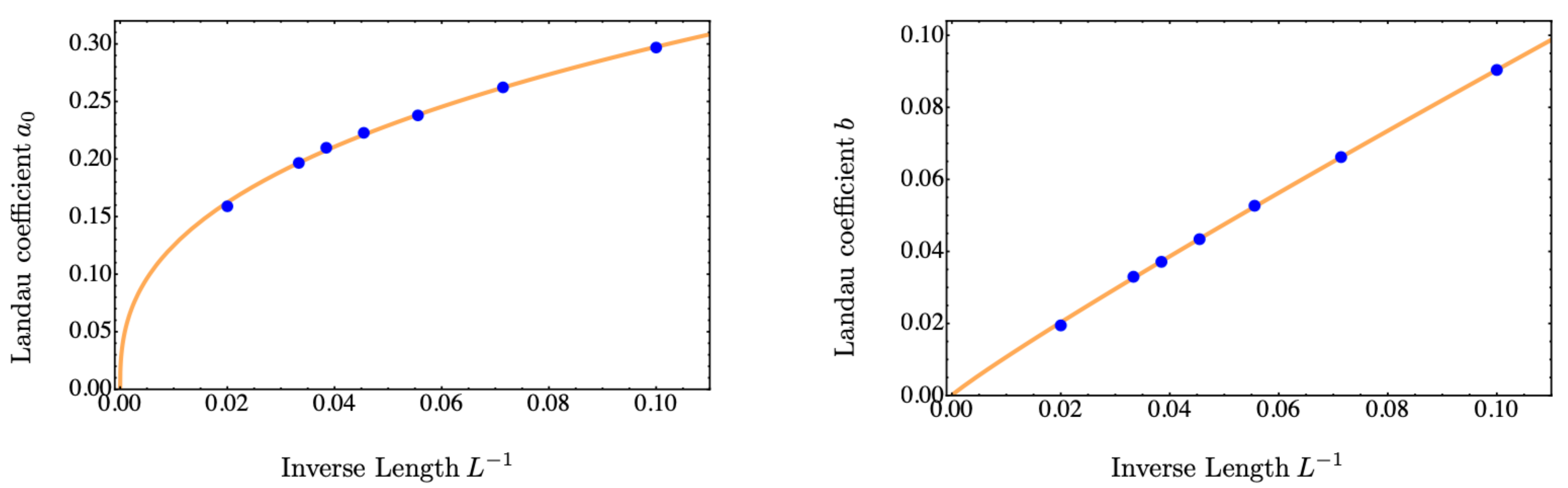}
\caption{Non linear fittings of Landau coefficients $a_0(L) = {\cal A} L^{(\gamma-1)/\nu}$ and $b(L) = \mathcal B L^{(\gamma-2\beta)/\nu}$, shown with the orange curves. The dots are the results of the fittings for $L = 10, 14, 22, 26, 30, 50$ given in SM.}
\label{a0bvsL}
\end{figure}

Next, since finite-size scaling theory indicates that the free energy should be of the form 
\begin{equation}
f(m,T;L) \approx L^{-(2-\alpha)/\nu} {\cal F}(tL^{1/\nu},m L^{\beta/\nu}) \>,\label{fFSS}
\end{equation}
with $t =|T-T_c|/T_c$, $2 - \alpha = \gamma + 2\beta$, and ${\cal F}(X,Y)$ an analytic function of its variables, then, in order to match this form, one finds that the mean field coefficients should scale as $a_0(L) \approx  {\cal A} L^{-(\gamma-1)/\nu}$ and $b(L) \approx {\cal B} L^{-(\gamma-2\beta)/\nu}$, with ${\cal A}$ and ${\cal B}$ non-universal amplitudes, 
again ignoring corrections to scaling for the moment. In this way, the mean-field free energy (\ref{LandauF}) can be written as,
\begin{widetext}
\begin{equation}
f(m,T;L) \approx f_0(T;L) +  L^{-(\gamma+2\beta)/\nu}  \left[  {\cal A} \>\left(m L^{\beta/\nu}\right)^2 \left(- {\cal L}_0+ L^{1/\nu}(T-T_c) \right) + {\cal B} \left(m L^{\beta/\nu}\right)^4 + \dots \right] \>,
\end{equation}
\end{widetext}
which, save for the unknown function $f_0(T;L)$, has the predicted finite-size scaling form given in (\ref{fFSS}). Fig. \ref{a0bvsL} shows the proposed fitting of $a_0(L)$ and $b(L)$ finding $\gamma = 1.237(1)$ and $\beta = 0.3263(5)$, again in very good agreement with the Ising exponents \cite{Butera:2002aa,Deng:2003aa,Lundow:2010aa,Ferrenberg2018,Chang:2025aa}, with the coefficients ${\cal A} = 0.7075(21)$ and ${\cal B}= 0.7647(22) $. One can further calculate the magnetic field $H$, see (\ref{SH});  setting $H = 0$ and considering $T < T_c^L$, the solution is the finite size spontaneous magnetization, but written in a finite-size scaling fashion,
\begin{equation}
m^2 \approx \frac{{\cal A}}{2{\cal B}} L^{-2\beta/\nu} \left({\cal L}_0 + L^{1/\nu}(T_c - T)\right) \>.\label{mFSS}
\end{equation}
This is also $m^2 \approx (a_0(L)/2b(L))(T_c^L-T)$, indicating the classical mean-field exponent $\beta_L = 1/2$ and, at the same time, the finite-size scaling form unveiling the Ising exponents $\nu$ and $\beta$. There is no discussion of this delicate point in the literature, to the best of our knowledge. Needless to say, an analogous conclusion regards the inverse susceptibility $\chi^{-1}$ showing both the classical  $\gamma_L = 1$ and the Ising $\gamma$ exponents.  In order for finite-size scaling to hold, higher order in Landau free energy must scale properly. This is certainly the case for the term $b^\prime(L) \approx {\cal B}^\prime L^{-(\gamma-2\beta-1)/\nu}$ shown in the Appendix. And, indeed, this coefficient grows as $L$ increases, since $\gamma-2\beta-1< 0$, thus reducing the region of validity of Landau free energy.\\

\begin{figure}[ht!]
\includegraphics[width=0.8\textwidth]{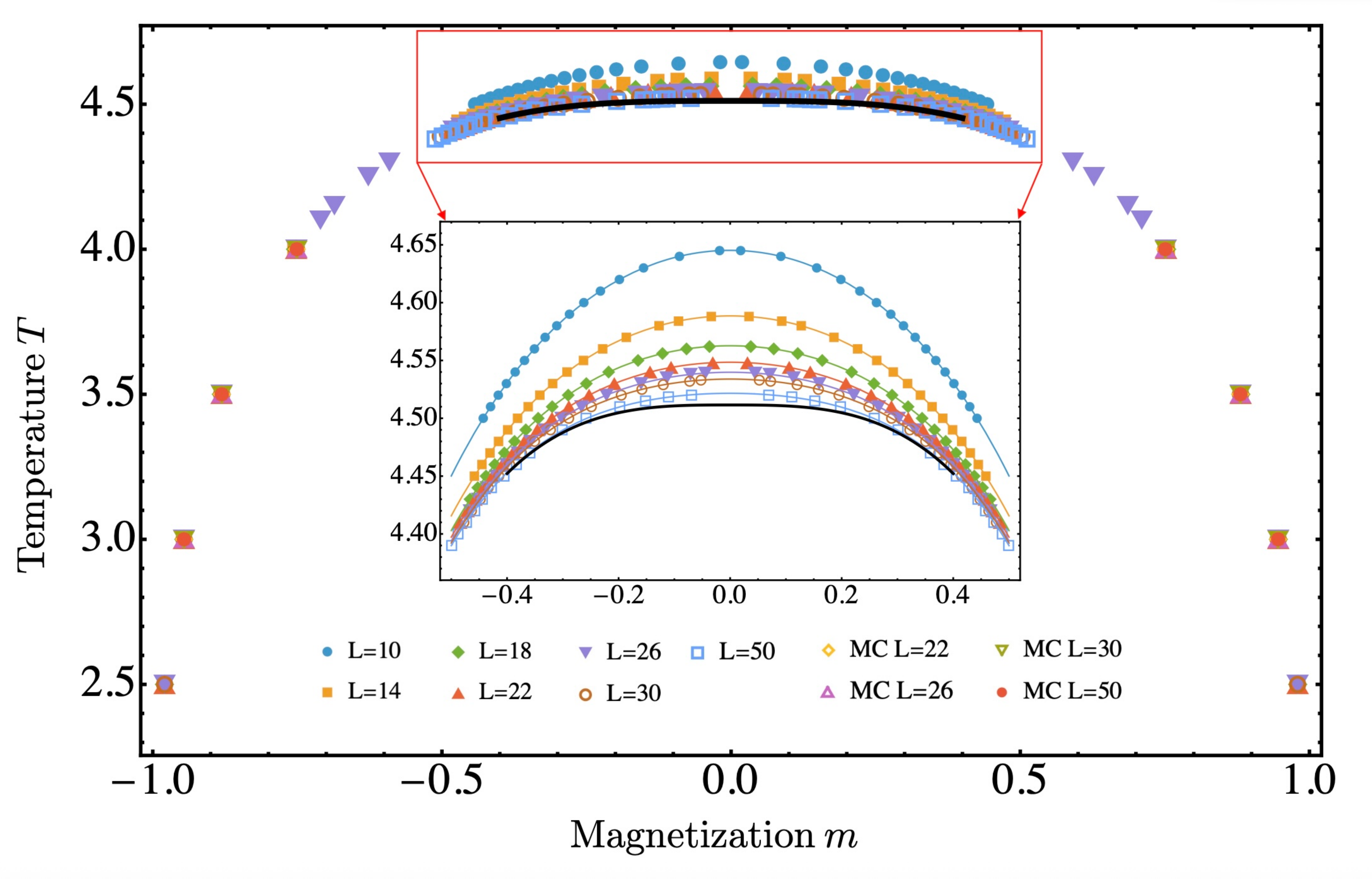}
\caption{Coexistence curves, or spontaneous magnetization, $T$ vs $m$, at $H = 0$, for systems with our studied values of  $L$. We also include values from standard MonteCarlo simulations marked as MC, for $T \le 4$, where
all points coalesce to the thermodynamic limit curve.  The inset shows  the spontaneous magnetizations near their critical points, each with an even polynomial fitting. 
The solid black curve is an empirical fitting of the coexistence curve in the thermodynamic limit. See text}
\label{mag}
\end{figure}

{\it The spontaneous magnetization--}As mentioned, we can extract the spontaneous magnetization, or coexistence curve shown in Fig. \ref{mag}, simply by finding the two minima of the free energy $f(m,T;L)$ for all values of $T < T_c^L$. Each coexistence curve, near $T_c^L$, can be fitted by an even polynomial, $T \approx T_c^L - C_2(L) m^2 - C_4(L) m^4 - C_6(L) m^6$, see the inset of Fig. \ref{mag}, finding,  within very small errors of few parts in $10^6$, the same values of $T_c^L$ as obtained with the fittings of the free energy. Further, very close to $T_c^L$, $T$ is quadratic in $m$, thus, $m^2 \approx (T_c^L-T)/C_2(L)$,  yielding again the exponent $\beta_L = 1/2$, with the  amplitude numerically agreeing with that obtained from the free energy approach, namely, $C_2(L) \approx 2 b(L)/a_0(L)$, and thus achieving the finite-size scaling form (\ref{mFSS}).\\

Additional insight into the transition from the mean-field behavior to the Ising one can be achieved by analyzing the full coexistence curve, without limiting ourselves to the critical region. Two aspects are important to highlight. One, the observation that {\it far} from the critical region, say $T \lesssim 4.0$, all curves appear to rapidly converge to the thermodynamic limit curve, as expected and observed in all sorts of simulations of coexisting phases \cite{Panagiotopoulos}. To further strengthen this statement, in Fig. \ref{mag} we have also included spontaneous magnetizations calculated with standard Monte Carlo simulations for temperatures far from the critical region, showing perfect agreement with the calculations of the present method for all values of $L$, incidentally further supporting its results. Second, shown with the solid black curve, we have included the accurate empirically calculated spontaneous magnetization for a cubic lattice, $m(t) = t^{0.32694109} (1.6919045 - 0.34357731 t^{0.50842026} - 0.42572366 t)$, see Refs. \cite{Liu:1989aa,Talapov:1996aa,Kaya:2022aa,Tambas:2023aa} for the different methods. Evidently, the sequence of coexistence curves appears to approach the true non-analytic thermodynamic limit curve as $L \to \infty$, while remaining mean-field in its critical vicinity as long as $L$ remains finite.
This indicates that the thermodynamic limit curve becomes so, from far to the critical region towards it, within a shrinking mean-field neighborhood.  In hindsight, the mean field behavior should have been expected in the light of the Ginzburg criterion \cite{Huang,Ma,Kardar}, though used differently. That is, close to the critical temperature, the size of the correlation length $\xi$ gets arrested at the size $L$ of the system, preventing it to develop its full critical behavior for closer temperatures and, therefore, it turns into the mean-field critical one with its own critical temperature $T_c^L$. In this way, the larger $L$ the smaller the region where Landau theory is valid. This subject must be further scrutinized.

{\it Final Remarks--}Although, due to limitations on computer time, we have here addressed the 3D Ising model only, we do have already verified that the finite-size free energy of the 2D Ising model shows the behavior exemplified in Fig. \ref{fvsm22}. Thus, we see no a priori objections for our findings here to be extended to $d = 2$ and $d \ge 4$. Nonetheless, this is a pending task to perform.
To conclude, we point out that while the purpose of the paper is not to find accurate exponents, as obtained with other well known and sophisticated techniques \cite{Butera:2002aa,Deng:2003aa,Lundow:2010aa,Ferrenberg2018,Chang:2025aa}, the method here used could certainly be optimized to obtain better values. For the present study the number of configurations was nearly ${\cal N} \sim N \times 10^{11}$ for every single temperature and system size. The present method can be extended to other systems, such as fluids with, for instance, the procedure developed in Ref.\cite{SastreIsing}.\\

{\it Acknowledgments--}The authors are grateful for financial support of PAPIIT-UNAM IN-108620, SECIHTI and Miztli Supercomputer Grant LANCAD-UNAM-DGTIC-082.

{\it Data Availability--}Data are available from the authors upon reasonable request.

\bibliography{citations}

\newpage
\appendix*
\section{\bf Appendix}

\begin{figure}[ht!]
\includegraphics[width=0.8\textwidth]{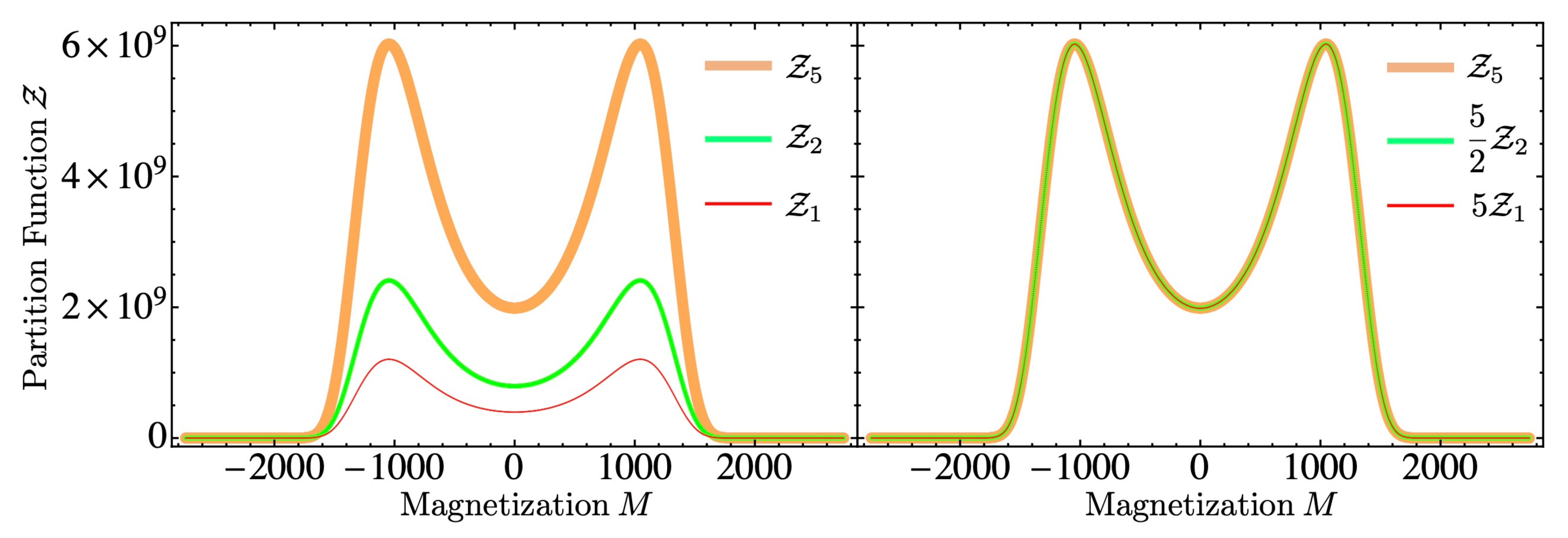}
\caption{Partition function ${\cal Z}_{\cal N}(N,T,M)$ vs magnetization $M$, for $L = 14$, at temperature $T = 4.50$. The left panel show ${\cal Z}_1$ for a given number ${\cal N}$ of configurations;  ${\cal Z}_2$ for $2{\cal N}$ and ${\cal Z}_5$ for $5{\cal N}$. The right panel shows the same partition functions coinciding, after being multiplied by ${5\cal Z}_1$, $(5/2){\cal Z}_2$ and ${\cal Z}_5$. This demonstrates that ${\cal Z}_{\cal N}(N,T,M)/{\cal Z}_{\cal N}(N,T,M^\prime)$ is independent of ${\cal N}$, depending only on $M$ and $M^\prime$, for given $N$ and $T$. In this calculation ${\cal N} = 14^3 \times 2 \times 10^8$.}
\label{compara}
\end{figure}

{\it Free energy calculation via modified Monte Carlo simulations--}As discussed in the main text, for a given value of the temperature $T$, number of spins $N = L^3$, and for a large sequence ${\cal N}$ of Monte Carlo trials, we find ${\cal Z}_{\cal N}(N,T,M)$ which is the number of configurations or spin states $\{s\} = (s_1,s_2, \dots , s_N)$ in the sequence, whose magnetization is $M_{\{s\}} = M$. This quantity is a fraction of the true partition function $Z(N,T,M)$. We note that all configurations in the sequence are counted, whether the spin flip is accepted or not; this indicates that the same state can be counted more than once. The main claim is expression (4) of the text that can be translated into the fact that the ratio ${\cal Z}_{\cal N}(N,T,M)/{\cal Z}_{\cal N}(N,T,M^\prime)$ is independent of ${\cal N}$ and depends on $M$ and $M^\prime$ only, for given $N$ and $T$. This is illustrated in Fig. \ref{compara}, where we show typical calculations of ${\cal Z}_{\cal N}(N,T,M)$ for the same $N$ and $T$ but different values of configurations, say ${\cal N}_1={\cal N}$, ${\cal N}_2=2{\cal N}$ and ${\cal N}_5=5{\cal N}$. One finds that, for values close to $M = 0$, ${Z}_{{\cal N}_5}(N,T,M) \approx (5/2) {Z}_{{\cal N}_2}(N,T,M) \approx 5 {Z}_{{\cal N}_1}(N,T,M)$.
Evidently, as the number of configurations is increased, ${\cal Z}_{\cal N}(N,T,M)$ includes more less probable magnetization states, which are those with $M \to \pm N$. This indicates that for large ${\cal N}$, ${\cal Z}_{\cal N}(N,T,M) \approx \zeta({\cal N},T,N) Z(M,T,N)$, where $Z(M,T,N)$ is the true partition function, and with $\zeta({\cal N},T,N)$ independent of $M$ and linear in ${\cal N}$ but still depending on $N$ and $T$. The free energy that we calculate is $f(m,T;L) = -(T/N) \ln {\cal Z}_{\cal N}(N,T,M) \approx -(T/N) \ln Z(N,T,M) - (T/N) \ln \zeta({\cal N},T,N)$ with $m = M/N$.  Its only unknown quantity is the free energy evaluated at $M = 0$, namely, $f(0,T;L) \approx -(T/N) \ln Z(N,T,0) - (T/N) \ln \zeta({\cal N},T,N)$, but derivatives with respect to $M$ can be calculated. The fact that the factor $\zeta({\cal N},T,N)$ depends on ${\cal N}$ and $T$ can be verified by calculating $f(m,T;L)$ for the same $L$ and ${\cal N}$ but different temperatures: one finds (not shown here) that the resulting $f(m,T;L)$ is not a concave function of $T$ for a given value of the magnetization $m$. To have the correct concavity, the relative number of configurations ${\cal N}$ should be larger the higher the temperature, but we cannot know this in advance, as this would be part of the solution we are seeking. The consequence of this is that quantities that depend on derivatives of the free energy with respect to $T$ cannot be calculated, such as the specific heat. Nevertheless, one still can calculate the magnetic field, the magnetic susceptibility and, more importantly, the spontaneous magnetization. Taking this minor drawback into consideration, we have performed calculations with essentially the same number of Monte Carlo trials ${\cal N}$ for a given $L$, using as a criterion the obtention of numerically accurate fits of $f(m,T;L)$ and, of course, our computer time limitations. In Table \ref{tabla1} we show the typical number of values of  ${\cal N}$ for different $N = L^3$.

\begin{table}[h]
    \centering
    \caption{Number of attempted spin flips for different size systems $L$, and for any single temperature $T$. The total number of sampled configurations is ${\cal N} = L^3 \times {\cal M} \times 10^9$. These calculations took of the order of 150,000 CPU hours in the {\it Miztli} supercomputer at UNAM.\\ }
    \label{tabla1}
    \begin{tabular}{lcccccccc}
        \hline\hline
        $L$ & 10 & 14 & 18 & 22 & 26 & 30 & 50 \\
         \hline
        $\cal{M}$& 2 & 5 & 4.5 & 5 & 4.25 & 2.75 & 1.8 \\
                \hline\hline
    \end{tabular}
\end{table}

{\it Analytic power expansion of the free energy, finite size scaling fittings and higher order corrections--}
The free energy $f(m,T,L)$, by symmetry considerations and verified by the calculation, is an even function of $m$, that can be written as equation (\ref{ffit}) of the main text.
%\begin{equation}
%f(m,T;L) = f_0(T,L) + A_2(T,L) m^2 + A_4(T,L) m^4 + A_6(T,L) m^6 +  A_8(T,L) m^8 + \dots \>. \label{ffit}
%\end{equation}
Again, the correct value of $f_0(T,L)$ is unknown, but, for a given value of $T$, this is just a shift, therefore, the values of $A_{2n}(T,L)$ can be correctly extracted. As mentioned in the text, the numerical evidence is that there exists a temperature $T_c^L$ at which $A_2(T_c^L,L) = 0$. Since the system is finite, $f(m,T;L)$ should be analytic at $T_c^L$ and, thus, the coefficients should have a power series in $(T-T_c^L)$, that is $A_{2n}(T,L) \approx A_{2n}(T_c^L,L) + A^\prime_{2n}(T_c^L,L)(T-T_c^L) + \cdots$. Our calculations allow for finding very precise values of $A_2(T,L) \approx a_0(L)(T-T_c^L)$ hence finding $T_c^L$ and $a_0(L)$ and $A_4(T,L) \approx b(L) + b^\prime(L)(T-T_c^L)$ but still statistically poor for obtaining $A_6(L,T_c) \approx c(L)$.  The obtained coefficients are given in Table \ref{tabla2}. 
\begin{table}[h]
    \centering
    \caption{Landau free energy critical temperature $T_c^L$ and coefficients $a_0(L)$, $b(L)$ and $b^\prime(L)$ for different size systems $L$, see equation (1) of the main text. \\ }
    \label{tabla2}
    \begin{tabular}{lcccccccc}
        \hline\hline
        $L$ & &$T_c^L$ && $a_0(L)$ && $b(L)$  && $b^\prime(L)$ \\
        \hline
        10
        &
        & 4.644938(32)
        &
        & 0.29686(60)
        &
        & 0.09038(11) 
        &
        & 0.1613(69)
        \\
       14
       &
        & 4.588280(62)
        &
        & 0.26222(86)
        &
        & 0.066191(38) 
        &
        & 0.1776(20)
        \\
        18
        &
        & 4.562497(69)
        &
        & 0.2379(11)
        &
        & 0.052683(61)
        &
        & 0.2304(41)
        \\
        22
        &
        & 4.548292(83) 
        &
        & 0.2228(12)
        &
        & 0.04340(19)
        &
        & 0.254(12)
        \\
       26
       &
        & 4.539717(31)
        &
        &0.20979(65)
        &
        & 0.03711(23)
        &
        & 0.304(23)
        \\
        30
        &
        & 4.533996(46)
        &
        & 0.1966(11)
        &
        & 0.03296(13)
        &
        & 0.328(17)
        \\
        50
        &
        & 4.521138(55)
        &
        & 0.1589(11)
        &
        & 0.01945(45)
        &
        & 0.440(57)
        \\
        \hline\hline
    \end{tabular}
\end{table}

\begin{figure}[ht!]
\includegraphics[width=0.5\textwidth]{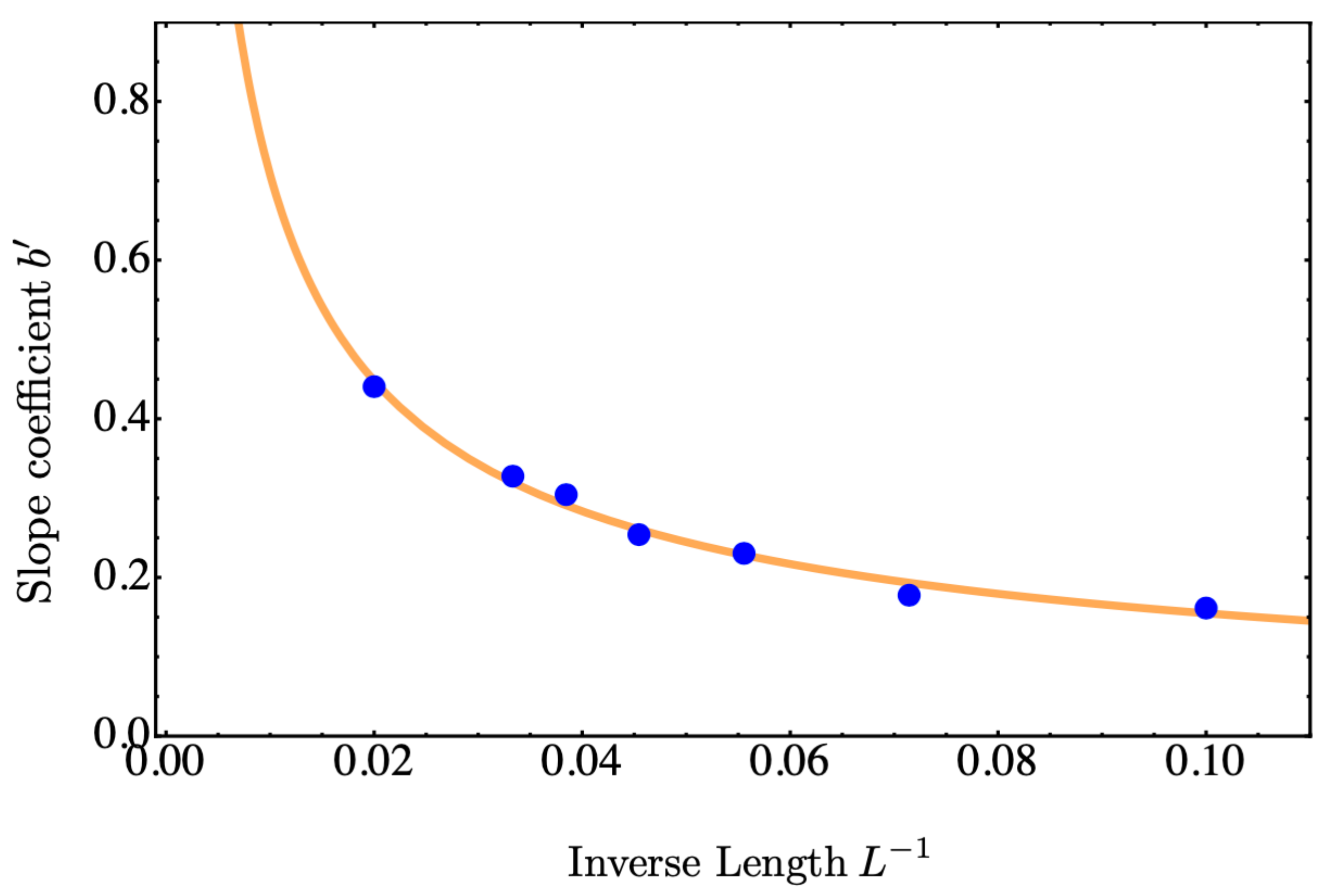}
\caption{Non linear fittings of the slope coefficient $b^\prime(L) = {\cal B}^\prime L^{-(\gamma-2\beta-1)/\nu}$ of the fourth order coefficient $A_4(T,L) \approx b(L) + b^\prime(L)(T-T_c^L)$ shown with the orange curves. The dots are the results of the fittings for $L = 10, 14, 22, 26, 30, 50$ given in Table \ref{tabla2}.}
\label{bprime}
\end{figure}

In Figs. 3 and 4 of the text we show the finite size scaling of Landau coefficients $T_c^L$, $a_0(L)$ and $b(L)$, however, an important task to verify such a scaling is the analysis of the next higher order terms in the Landau expansion. As discussed in the text, these are $b^\prime(L) \approx {\cal B}^\prime L^{-(\gamma-2\beta-1)/\nu}$ and $c(L) \approx {\cal C} L^{-(\gamma-4\beta)/\nu}$. While our calculations are not enough to obtain the latter, we can, at least, find the former. This is shown in Fig. \ref{bprime}, where we use the obtained values of the exponents $\beta$ and $\gamma$ and fit the coefficient $\mathcal B$ only; the fit is very satisfying indicating that finite size scaling indeed is obeyed at this higher order. The fact that $b^\prime(L)$ diverges as $L$ grows signals that the region where Landau theory is valid, becomes smaller for larger values of $L$.

\end{document}